\documentclass[aps,prl,twocolumn,floatfix]{revtex4}
\usepackage{graphicx}

\begin{document}
\draft
\title{Measured oscillations of the velocity and temperature fields
in turbulent Rayleigh-B\'{e}nard convection in a rectangular cell}
\author{Sheng-Qi Zhou, Chao Sun and Ke-Qing Xia}
\address{Department of Physics, The Chinese University of Hong Kong, Satin,
Hong Kong, China}
\date{\today}

\begin{abstract}
Temperature and velocity oscillations have been found in a
rectangular Rayleigh-B\'{e}nard convection cell, in which one large
scale convection roll exists. At $Ra=8.9 \times 10^{11}$ and $Pr=4$,
temperature oscillation can be observed in most part of the system
and the oscillation period remains almost constant, $t_T = 74 \pm 2$
seconds. Velocity oscillation can only be found in its horizontal
component, $v_y$ (perpendicular to large scale circulation plane),
near the cell sidewall, its oscillation period is also constant,
$t_T = 65 \pm 2$ seconds, at these positions. Temperature and
velocity oscillations have different Ra dependence, which are
respectively indicated by Peclect number $Pe_T = 0.55Ra^{0.47}$ and
$Pe_v = 0.28Ra^{0.50}$ . In comparison to the case of cylindrical
cell, we find that the velocity oscillation is affected by the
system geometry.
%To illustrate the influence of system geometry on convective flow,
%these results have been discussed and compared to the previous
%observations in the cylindrical case.
%The behavior of temperature oscillation here is found to be
%consistent with that in the cylindrical cell. The velocity one shows
%different scalings in the cells of different geometries, but the
%velocity oscillation time exactly agrees with its circulation time
%of large scale flow in the same system.
\end{abstract}
\bigskip
\narrowtext

\maketitle

\section{INTRODUCTION}
The coherent large-scale flow structure is one of the most important
constituents in  turbulent thermal convections, which occur widely
in the oceans, the atmosphere, the earth's outer core and galaxies
\cite{Atkinson96}. Its detailed flow dynamics has been mainly
investigated in the classical Rayleigh-B{\'e}nard convection system,
where a confined enclosure is heated from below and cooled from
above \cite{MALKUS54,Howard64,Siggia94}. The system has three
control parameters, namely the Rayleigh number ($Ra$), Prandtl
number ($Pr$) and the aspect ratio ($\Gamma$). The Rayleigh number
is defined as $Ra=\alpha g L^3 \Delta T/\nu \kappa$, with $g$ being
the gravitational acceleration, and $\alpha$, $\nu$, and $\kappa$
being, respectively, the thermal expansion coefficient, the
kinematic viscosity, and the thermal diffusivity of working fluid.
The Prandtl number is defined as Pr$=\nu / \kappa$, and the aspect
ratio $\Gamma$ is the ratio of the lateral dimension of the system
to its height. In the system, a prominent self-sustained low
frequency oscillation of the measured local velocity and/or
temperature has been generally observed in numerous experiments and
numerical simulations
\cite{Howard64,Siggia94,Krishnamurti81,Heslot87, Sano89, Castaing89,
Villermaux95, Takeshita96, Ciliberto96, Cioni97, Ashkenazi99,
Krishnamurti99,Niemela01, Shang01, Qiu01n, Qiu01,Qiu02, Qiu04,
Shang04, Funfschilling04, Sun05,Sun05a, Resagk06,
Verdoold06,Puits07}. However, the origin of this characteristic
frequency still remains elusive although much effort has been made
to tackle this issue in the past decades.

Explanation of low frequency oscillations can be traced to the
``bubble model" \cite{Howard64} and later the ``plume model"
\cite{Krishnamurti99}, which were developed based on the presence
and persistence of thermal structures ({\it{e.g.}} thermals, plumes)
\cite{MALKUS54}. It is assumed that a thermal boundary layer
thickens in time by the upward diffusion of the temperature field.
When a critical thickness is attained, the boundary layer will erupt
and quickly expel all the buoyant fluid through plumes and thermals
away from the boundary. The depleted boundary layer thus grows once
more by diffusion and the process repeats in a periodic manner. Such
a phenomenological picture has been observed in high Pr fluid with
the shadowgraph technique \cite{Xi04}. In later experiments, it has
been observed that the unique frequency is related to the
large-scale circulation (LSC) in a helium gas cylindrical cell with
$\Gamma=1$ \cite{Heslot87,Sano89,Castaing89, Qiu04}, and its value
is more or less the same as the circulation frequency of the large
convective flow. A coupled-oscillator model was proposed to explain
the observed phenomenon, in which the plume clusters from one plate
are triggered by those from the opposite plate and the overall
process will be self-sustained \cite{Villermaux95}. A prediction of
the model is that the oscillation period is twice of the
cell-crossing time of the clustered plumes. This has been verified
by later direct velocity measurements \cite{Qiu01n,Qiu01,Sun05}. All
the above mentioned models suggest that the thermal plumes and the
thermal boundary layers play a dominant role in the generation of
the oscillation.

In thermal convection, it has been found that, for the most part,
the LSC has a preferred orientation in $\Gamma=1$ cylindrical cells
and it oscillates azimuthally around this orientation \cite{Cioni97,
Funfschilling04,Sun05,Brown06,Xi06}. Recently, several new dynamic
processes, {\it{e.g.}} reorientation, cessation and so on, have been
observed in the cylindrical cells \cite{Niemela01,Brown06,Xi06}, but
it has been argued that such processes have little relevance to the
azimuthal oscillation \cite{Fontenele05}.  Based on the phenomenon
of horizontal oscillations of LSC \cite{Funfschilling04}, a new
dynamical model has been proposed \cite{Resagk06,Puits07}. This
model suggests that the horizontal oscillation of large scale flow
is the main source of the low frequency oscillation in the system.
Instead of being caused by the periodic plume-ejections from the
boundary layers, this horizontal oscillation is intrinsic to the
bulk dynamics. The periodical ejections of plumes may then be viewed
as passive objects of the process.

The LSC dynamical picture can be simplified if the large scale
reorientations or cessations are not considered, the horizontal
oscillation of LSC will be analogous to a classical harmonic
oscillator wandering around its origin. In this case, one may wonder
whether the system geometry plays a role in the horizontal
oscillations of the LSC. In most cases experiments and simulations
have been performed mainly in a cylindrical cell of aspect ratio one
in which axis-symmetry holds and only one large scale convective
roll exists
\cite{Heslot87,Sano89,Castaing89,Villermaux95,Takeshita96,
Krishnamurti99,Niemela01,Shang01, Qiu01n, Qiu01,Qiu02, Qiu04,
Shang04, Funfschilling04,Sun05,Sun05a, Xi04,Brown06,Xi06,Puits07}.
At low Ra, the linear stability analysis predicted that one
asymmetric convection roll occupies the whole bulk regime when the
cell provides an axis-symmetric boundary condition
\cite{Charlson71}, and the convection roll could be aligned in any
direction around the circle \cite{Golubitsky83}. But in real systems
there are inevitably imperfections, initial instabilities and other
external factors, which lead to the evolution of one stable branch,
with all other branches being unstable \cite{Neumann90}. In
numerical simulations, similar picture could be expected for the
case of high Ra convection \cite{Verzicco97}.

In this paper, we report turbulent thermal convection experiments in
a rectangular cell, in which the axis-symmetry is absent. Local
velocity and temperature measurements at various locations were made
over varying values of Ra and at a fixed Pr. We focus on the
oscillatory behavior of measured local velocity and temperature
fields. The remaining sections of the paper are organized as
follows. In Sec. II we describe the rectangular convection cell and
the velocity and temperature measurements. In Sec. III, we present
and discuss the experimental results. There we first show the
atuo-correlation functions and histograms of the velocity and
temperature signals. We also discuss the spatial distribution of the
velocity and temperature oscillations. We then present the
Ra-dependence of the periods of these oscillations. A brief summary
is given in Sec. IV.

\section{Experiment setup}
\begin{figure}
\includegraphics[width=2.5in]{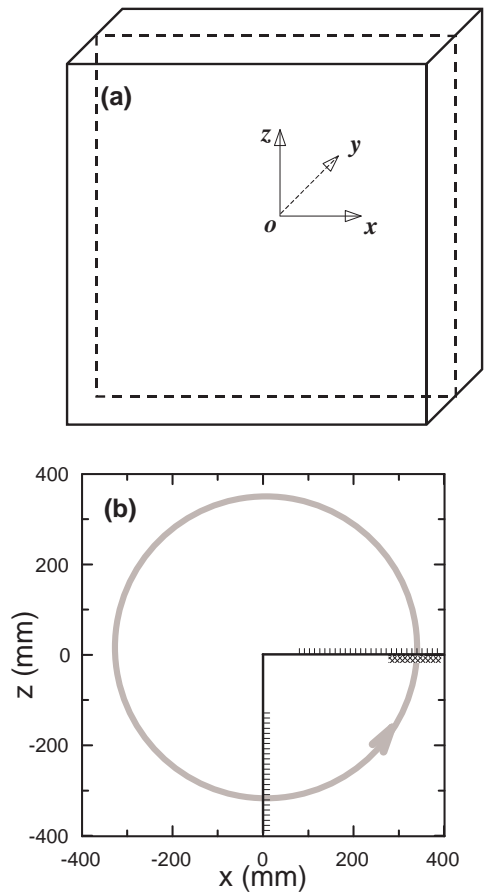}
\caption{(a) Schematic drawing of the convection cell and the
coordinates for the experiment; (b) The $x-z$ cross section plane of
the cell at $y=0$ (shown as the dashed plane in (a)). Inside the
plane, the large scale flow is shown as the gray circle. The
measured positions are along the marked solid lines, the positions
where the oscillation can be found from temperature signal are
indicated by the line-combs, those from velocity by the
line-diamonds.} \label{fig:setup}
\end{figure}

 The convection cell has a square vertical cross section, its
length, width, and height are $81\times20\times81$ (cm), as shown in
Fig.~\ref{fig:setup} (a). With this geometry, the axis-symmetry does
not exist, and the large-scale flow is expected to be largely
confined in the plane with the aspect ratio $\Gamma=1$. The working
fluid was water. Detailed description of the apparatus can be found
in Ref.\cite{Xia03}.  Here the coordinates are defined as follows:
the origin coincides with the cell center, its $x$ axis points to
the right, the $z$ axis points upward, and the $y$ axis points
inward. We measured the velocity and temperature signals at
positions marked by the solid lines in Fig.~\ref{fig:setup} (b). The
velocity was measured by Laser Doppler Anemometry (LDA), the
measurement volume was approximately 0.12 mm in diameter and 1.1 mm
in length \cite{Shang01}. Along the line ($x,0,0$), two-dimensional
velocity ($v_y$, $v_z$) was measured, and velocity ($v_x$, $v_z$)
was measured along the line ($0,0,z$). At each position $\sim 10^6$
or at least $5.2$ hrs statistically independent samples were
acquired. The temperature measurement apparatus was similar to that
described in Ref.\cite{Zhou01n}. The local temperature was measured
by using a thermistor about $300$ $\mu$m in size. The thermistor
served as one arm of an AC Wheatstone bridge driven sinusoidally at
1kHz in frequency and $0.5$ V in amplitude. The output of the bridge
was first fed to a lock-in amplifier and then digitized by a dynamic
signal analyzer at a sampling rate of $128$ Hz, the recording time
was $4.5$ hrs to ensure good data statistics. Ra and Pr was kept at
$8.9 \times 10^{11}$ and $4.0$ respectively
 unless stated otherwise.

\section{Results and discussion}
\subsection{Autocorrelation function and histogram}
\begin{figure}
\includegraphics[width=3.5in]{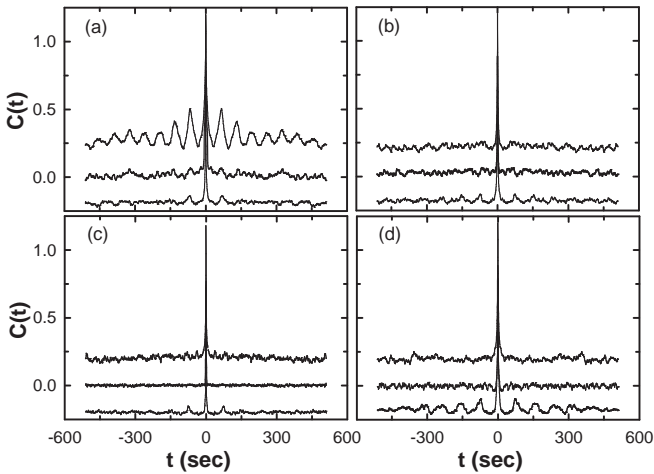}
\caption{(a) Top to bottom curves: auto-correlation functions of
$v_y$ (shifted up by $0.2$), $v_z$ and temperature (shifted down by
$0.2$) at ($385$ mm, $0,0$); (b) same as (a) but at ($240$ mm,
$0,0$); (c) top to bottom curves: auto-correlation functions of
$v_x$ (shifted up by $0.2$), $v_z$ and temperature (shifted down by
$0.2$) at ($0,0$, $-385$ mm); (d) same as (c) but at ($0,0$, $-240$
mm).} \label{fig:corr}
\end{figure}

To test whether a low frequency oscillation exists we analyze the
auto-correlation function, $C(t)=<{\delta}X(\tau){\delta}X(\tau+t)>$
of these measured signals, here ${\delta}X$ is the fluctuation of
$X$ normalized by its RMS value, $X$ can be $v_x$, $v_y$, $v_z$ and
$T$ and $<>$ denotes the time average. Examples of the
auto-correlations of $v_y$, $v_z$ and $T$ at ($385$ mm,0,0) near the
sidewall are shown in Fig.~\ref{fig:corr} (a). An oscillation can be
found in $v_y$, whose intensity, represented by the magnitude of the
second order peak, is as strong as that observed in the vertical
velocity measurement near the sidewall of cylindrical cells
\cite{Qiu04,Shang04,Shang01}. But the vertical velocity $v_z$ here
shows no oscillation from its auto-correlation function, which
differs from that in the cylindrical cells
\cite{Qiu04,Shang04,Shang01}. The oscillation of temperature signal
is much weaker than that in the cylindrical cell \cite{Shang04}, but
the second order peak can be well resolved from the background noise
fluctuation. The oscillation periods obtained from $v_y$ and $T$ are
denoted as $t_v$ and $t_T$ respectively. When the measurement moves
away from the sidewall to the region of the cell, as shown in
Fig.~\ref{fig:corr} (b) where the location is ($240$ mm,$0,0$), the
velocity oscillation can not be observed from both $v_y$ and $v_z$
components, whereas temperature oscillation can still be detected.
In the measurement along the line ($0,0,z$), two examples at ($0,0$,
$-385$ mm) and ($0,0$, $-240$ mm) are shown in Fig.~\ref{fig:corr}
(c) and (d) respectively, in which only temperature oscillation can
be observed, the velocity one cannot be detected in either $v_x$ or
$v_z$ component.

\begin{figure}
\includegraphics[width=3.5in]{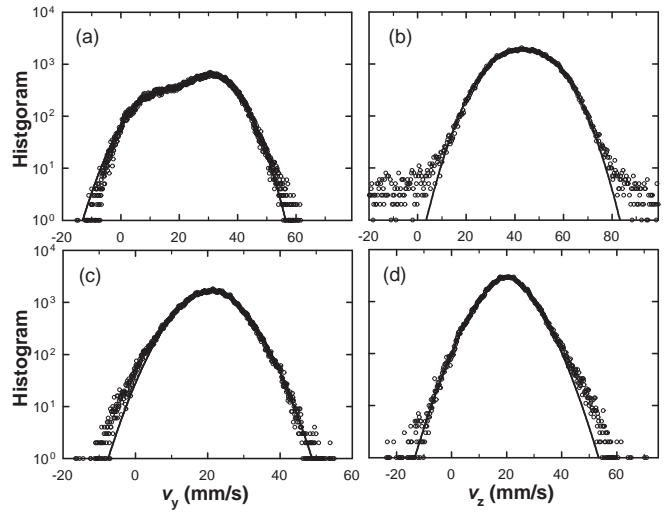}
\caption{Histograms of (a) $v_y$ at ($385$ mm, $0,0$), (b) $v_z$ at
($385$ mm, $0,0$), (c) $v_y$ at ($240$ mm, $0,0$) and (d) $v_z$ at
($240$ mm, $0,0$). Their fitting curves are plotted in solid lines
from (a) two-peak Gaussian function, (b) one-peak Gaussian function,
(c) and (d) one-peak exponential power function. } \label{fig:pdf}
\end{figure}

We now examine the histograms of these velocity signals, some
examples are shown in Fig.~\ref{fig:pdf}. Along the line
($x,0,0$), there are two peaks in the histogram of $v_y$ at
location ($385$ mm, $0,0$) (Fig.~\ref{fig:pdf} (a)), and it can be
fitted with a two-peak Gaussian function. The distribution of
$v_y$ has two peaks which is consistent with the strong horizontal
velocity oscillation shown in its auto correlation function
(Fig.~\ref{fig:corr}(a)). The histogram of $v_z$
(Fig.~\ref{fig:pdf}(b)) exhibits typical Gaussian-like
distribution, it is also consistent with the property of its
auto-correlation function (Fig.~\ref{fig:corr}(a)). At location
($240$ mm, $0,0$), both velocity components $v_y$ and $v_z$ have
similar single-peaked histograms (Fig.~\ref{fig:pdf}(c) and (d)),
they can be fitted by a stretched exponential function ($\sim
exp[-{\frac{(x-x_0)}{b}}^\beta]$), here the fitted $\beta$ is
$1.6\pm0.1$.  We also examined the histograms for both the
 $v_x$ and $v_z$ velocity components along the line ($0,0,z$), and find single-peaked
Gaussian-like distributions for positions close to the bottom plate
and stretched-exponential distribution for locations far away from
the plate. Such results are consistent with the observations that no
oscillation in their auto-correlation functions
(Fig.~\ref{fig:corr}(b), (c) and (d)).

It is a surprise that the oscillation in the velocity field cannot
be observed in its vertical component $v_z$. Similar finding has
also been reported in the PIV measurement of the present system
\cite{Xia03}, where no velocity oscillation has been seen in the
large scale circulation plane. Such an observation is different from
that in the cylindrical cell and does not support the ``plume model"
explanations \cite{Howard64, Krishnamurti99, Villermaux95,Qiu01n,
Qiu01,Qiu02, Xi04, Qiu04, Sun05}. If the thermal plumes are ejected
from the conducting plates in a periodical manner
\cite{Qiu02,Qiu04,Xi04}, the vertical velocity $v_z$ should contain
an oscillatory component, as the thermal plumes are driven by the
buoyancy force. The absence of this oscillation suggests that
temperature and velocity oscillations may have different origins and
be controlled by different mechanisms.
%{\it
%e.g.} turbulent noise \cite{Brown07}.

\subsection{Spatial distribution}
One can see that both temperature and velocity oscillations can be
observed in the present system (although not in all positions). The
positions where the oscillation can be found from $T$ and $v_y$ are
marked in Fig.~\ref{fig:setup}(b). The temperature oscillation can
be detected in the range $x \in \{80$ mm, $395$ mm$\}$ along the
line ($x,0,0$) and $z \in \{-130$ mm, $-387$ mm$\}$ along the line
($0,0,z$). There is no temperature oscillation in the central region
of the cell. We examine the data in the cylindrical cell of
Ref.\cite{Zhou01} and find the same result. However in the
cylindrical cell, it has been found that the temperature oscillation
near the sidewall is much stronger than that near the bottom plate
\cite{Qiu02}. Here we cannot distinguish where the temperature
oscillation is obviously stronger from the auto correlations, as
shown in Fig.~\ref{fig:corr}, .

\begin{figure}
\includegraphics[width=3.0in]{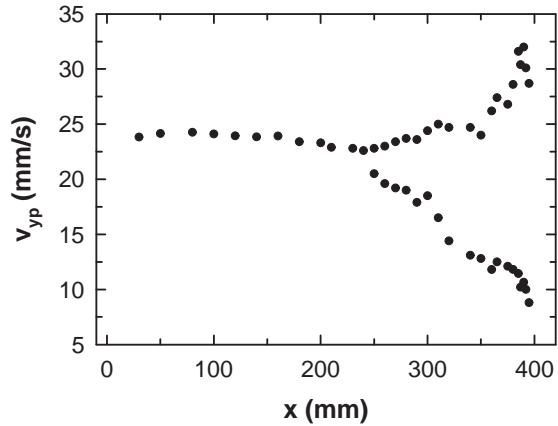}
\caption{The peak values of $v_y$ as function of $x$ when the
measured position moves from the cell center to the sidewall.}
\label{fig:angle}
\end{figure}

From Fig.~\ref{fig:setup}(b), we can see that velocity oscillation
can only be detected in its component $v_y$ near the sidewall ($x
\in \{250$ mm, $395$ mm$\}$). Correspondingly, in these positions
two peaks have been observed in the histogram of $v_y$. The values
of the peak positions in the histograms are shown in
Fig.~\ref{fig:angle}. The difference between the two peak values
becomes smaller as the measuring position moves away from the
sidewall to the central region, and the two values merge into one at
$x$ around $250$ mm, where the oscillation becomes undetectable in
the velocity field, as illustrated in Fig.~\ref{fig:setup}. No
oscillation can be found in other measured positions, including $x
\in \{0$, $250$ mm$\}$ along the line ($x,0,0$) and the whole line
($0,0,z$). The result is in sharp contrast to what is observed in
the cylindrical cell \cite{Qiu04,Sun05}.  In the cylindrical cell,
the oscillation could be found everywhere in the bulk regime from
the horizontal velocity measurement, and the oscillating intensity
is stronger in the center than that in other positions \cite{Sun05}.
Recently, the oscillation has also been explored in a rectangular
cell, where the horizontal cross section is square and the aspect
ratio $\Gamma=4$ \cite{Verdoold06}. The authors found that the
oscillation stems from the periodical switching between two
corotating convective rolls. Therefore, the oscillations observed in
the present system should not share the same dynamic mechanism as
that case because only one large scale roll exists here
\cite{Xia03}.
%In this work, a rectangular shaped cell was used and the working fluid was water.
 Unlike the recent
convection experiment in a rectangular box

\begin{figure}
\includegraphics[width=3.0in]{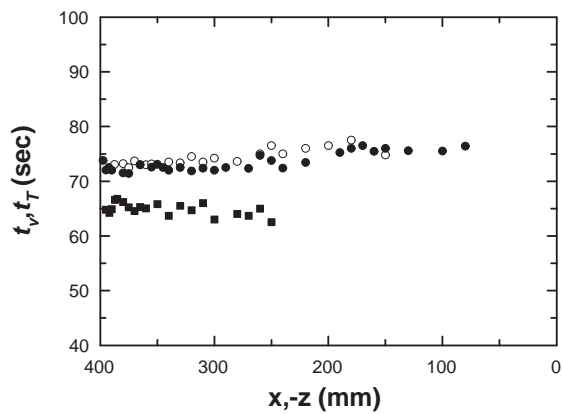}
\caption{The measured oscillation periods from the velocity $v_y$
along the line ($x,0,0$) (solid squares) and temperature signals
along the lines ($x,0,0$) (open circles) and ($0,0,z$) (solid
circles).} \label{fig:time}
\end{figure}

Next, we discuss the position dependence of the oscillation periods,
$t_T$ and $t_v$, obtained from temperature and velocity fields
repectively. As shown in Fig.~\ref{fig:time}, we can see that within
experimental uncertainty the temperature oscillation period ($t_T$)
may be regarded as a constant for all the positions measured, $t_T =
74 \pm 2$ seconds for both the vertical and horizontal scans. The
fact that temperature oscillation can be observed in most part of
the measured locations with the constant oscillation period implies
that it is a global characteristic of the convective flow and the
oscillation period is a typical time scale of the system.
Oscillation of the velocity field can only be observed in a rather
narrow region close to the sidewall, inside this region the
turbulent intensity was found to be the strongest in the system
\cite{Xia03}. The velocity oscillation period ($t_{v}$) also remains
more-or-less constant for all the measuring positions, $t_v = 65 \pm
2$ seconds. Note that $t_{v}$ is about $14\%$ less than $t_T$, such
a difference between the two oscillation periods is beyond the
experimental uncertainties for the respective measurements.

%It should be mentioned that
%a similar observation has been found in the azimuthal motion of LSC
%\cite{Xi06}, there is around $10\%$ difference between the
%oscillation periods of the orientation angle and magnitude of LSC.
%However, we are not aware how much relationship they have with the
%present finding.

\subsection{Ra dependence}

\begin{figure}
\includegraphics[width=3.0in]{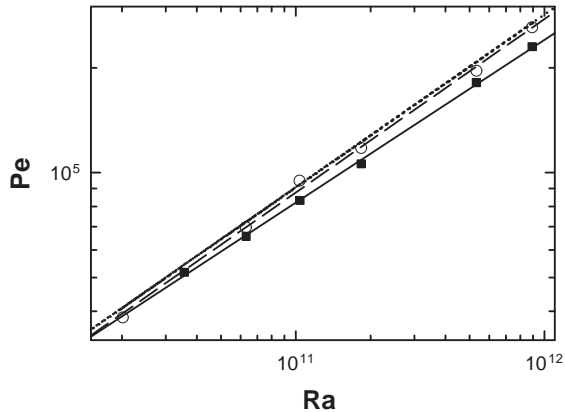}
\caption{The Ra-dependence of Peclet number. $Pe_v$ from the
velocity oscillation (open circles) and its fitting $0.29Ra^{0.50}$
is plotted as dash line; $Pe_T$ from the temperature oscillation
(solid squares) and its fitting $0.55Ra^{0.47}$ is plotted as solid
line; Peclet number obtained from the rotational rate in the PIV
measurement \cite{Xia03} is plotted here as dotted line.}
\label{fig:Ra}
\end{figure}

Keeping constant Pr$=4.0$, we also measured the oscillation periods
($t_{T}$ and $t_{v}$) of temperature and velocity fields over the
range of Ra from $2\times10^{10}$ to $8.9\times10^{11}$. In order to
compare to the results in previous literatures
\cite{Takeshita96,Niemela01,Qiu02,Qiu04,Funfschilling04,Xia03},
$t_{T(v)}$ is normalized in the form of the Peclet number
($Pe_{T(v)}\equiv\frac{UL}{\kappa}=\frac{4L^2}{t_{T(v)}\kappa}$),
here the subscript of $Pe$ denotes that it is obtained from either
the temperature or velocity, and the typical velocity U is
represented by $\frac{4L}{t_{T(v)}}$. As shown in Fig.~\ref{fig:Ra},
we can see that $Pe_T$ and $Pe_v$ have comparable magnitude, but
their difference can be well discerned. The Ra-dependence of $Pe_T$
can be fitted with $0.55Ra^{0.47}$. The exponent agrees excellently
with the temperature oscillation measurements in low temperature
helium gas \cite{Niemela01}, mercury \cite{Takeshita96}, and water
\cite{Qiu02} of a cylindrical cell, the different prefactors are
perhaps due to the fluids of  different Pr. Moreover, the scaling is
consistent with that of the angular oscillations of the horizontal
thermal plume motion above the bottom plate in a cylindrical cell
\cite{Funfschilling04}. This result confirms that the temperature
oscillation is the global and universal character in thermal
convection, and this result also implies that temperature
oscillation period ($t_T$) is a result of flow motions of thermal
structures and independent of system geometry. We also notice that
some experiments have reported different scalings for temperature
oscillation period ($t_T$) in the cylindrical cell
\cite{Castaing89,Cioni97,Sun05a}. One possible explanation is that
the pathlength of the circulation affects the scaling of oscillation
period \cite{Sun05a}.
%Recently, it was suggested that the
%pathlength of the circulation would affect the scaling of $Pe_T~Ra$
%in the cylindrical cell\cite{Sun05a}, it was said that the
%pathlength

In the cylindrical cell, it was reported that the velocity
oscillation follows $0.167 Ra^{0.47}$ and it is driven by the same
mechanism as that of temperature \cite{Qiu02,Qiu04}. That is, the
velocity and temperature oscillations are generated by the
alternating emission of thermal plumes between the upper and lower
thermal boundary layers \cite{Qiu04}. Moreover, it was found that
both temperature and velocity oscillation periods are coincident
with the rotational period of the large scale circulation
\cite{Qiu01,Qiu01n,Qiu04}. In the present system, the
$Ra$-dependence of $Pe_v$ is fitted as $0.29Ra^{0.50}$. In a
previous PIV measurement \cite{Xia03}, the rotation rate of large
scale circulation has been obtained, and the Pe could be fitted with
the formula, $0.318Ra^{0.496}$, which is re-plotted in
 Fig.~\ref{fig:Ra}. We can see the Peclet number $Pe_v$  agrees this
 fitting line very well, but the temperature $Pe_T$ does not. These
results suggest that the velocity oscillation period is consistent
with the rotational period of the large scale flow in both
cylindrical and rectangular cells, and both of them could be
affected by the system geometry. Based on the results that
temperature and velocity oscillations have different responses to
the system geometry, it may be proposed that the two oscillations
reflect different aspects of the thermal convective flow even though
they are related to each other.

\section{Conclusion}

In this work, both temperature and velocity oscillations can be
found in the rectangular cell, where only one large-scale convection
roll exists. At fixed Ra ($8.9 \times 10^{11}$), it is found that
temperature oscillation can be observed in most of the locations in
which measurements were made, and they gave a position-independent
oscillation period, $t_T = 74 \pm 2$ seconds. For measurements made
with different $Ra$, the normalized oscillation period $Pe_T$ is
found to scale as $0.55Ra^{0.47}$, which is the same as that found
in cylindrical cells \cite{Takeshita96,Niemela01,Qiu02,Qiu04}. This
scaling is also the same as that obtained from the horizontal
motions of thermal structures (plumes) \cite{Funfschilling04}. This
result suggests that temperature oscillation is a global and
universal characteristic of turbulent thermal convection, and its
statistical behavior may be related to the flow motions of thermal
structures and are independent of system geometry.

It is a surprise that the oscillation in the velocity field cannot
be seen in its vertical component $v_z$ near the sidewall. If the
thermal bursts are erupted in a periodical way, it should be
detected by the vertical velocity because the thermal bursts are
driven vertically under the buoyancy force. However, velocity
oscillation can only be observed in its horizontal component ($v_y$)
at positions close to the sidewall, where its histogram exhibits two
peaks. The normalized velocity oscillation period $Pe_v$ shows an
Ra-dependence of $Ra^{0.50}$ and it collapses very well on that
obtained from the rotation rate of convective flow \cite{Xia03}.
This is different from the observation in the cylindrical cell
\cite{Qiu04}, where it was reported that the periods for both the
velocity and temperature oscillations and the rotational period of
the convective flow follow the same scaling \cite{Qiu01n,Qiu04}.
This comparison suggests that the velocity oscillation period is
consistent with that obtained from rotation rate of the large scale
flow, and both of them could be affected by the system geometry.

We would like to thank G. Ahlers and H.-D. Xi for illuminating
discussions. K.-Q. Xia was supported by Hong Kong RGC Grant 403705.
S.-Q. Zhou was supported by CUHK direct grant 2060309 and United
College research grant CA11096.
% in qiu01's paper \gamma_u=u_{max}/(L/2)

\end{document}